\documentclass{article}
\usepackage{amssymb}
\usepackage{graphics}
\usepackage{epsfig }
\usepackage{subfigure}
\usepackage{graphicx}
\usepackage{amsmath}

\begin{document}
\begin{titlepage}

\title{Effective cosmological ``constant'' and
quintessence}
\author{R. de Ritis$^{1,2}$, A. A. Marino$^{1,3}$ \\
{\em \small $^1$ Dipartimento di Scienze Fisiche, Universit\`{a} di Napoli,} \\
{\em\small $^2$Istituto Nazionale di Fisica Nucleare, Sezione di Napoli,}\\
 {\em \small Complesso Universitario di Monte S. Angelo, Via Cinzia, Edificio N
 80126 Napoli, Italy,}\\
 {\em\small$^3$Osservatorio Astronomico di Capodimonte,}\\
{\em\small Via Moiariello, 16-80131 Napoli, Italy}}.

\maketitle
\date{}

\begin{abstract}
In this paper we present exactly solved extended quintessence models;
furthermore, through a dynamical effective Q-cosmological "constant", we
recover some of the $\Lambda$ decaying cases found in the literature.
Finally we introduce a sort of complementarity between the Q-dominated or
$\Lambda$-dominated expansions of the Universe.
\end{abstract}

\vspace{20.mm}
e-mail addresses:\\deritis@na.infn.it\\marino@na.astro.it\\
\vfill
\end{titlepage}
\noindent
There are  recent crucial informations coming from obeservations which seem
to put new light on our nowadays knowledge of the Universe. We mention just
two of them: first, the recent observations on Type Ia supernovae are
strongly in favour of an accelerated Universe \cite{perl1} \cite{perl2}
\cite{leib}. As well, measurements of microwave background, mass power
spectrum \cite{wang} and lensing statistics (for example, [5]) all suggest
that a large amount of energy density in the Universe should have a
negative pressure. That is, we have to include some missing part of the
energy needed to reach the critical one. Second, as suggested by BOOMERanG
\cite{bern}, it seems to be very reasonable to assume that the Universe is
spatially flat, that is $\Omega_{k}=0$ (in agreement with inflationary
scenarios) as we will consider  all along the paper. In connection with the
first information we will follow the assumption present in the literature
assuming that the missing part of the energy density can be supposed to
have the form of "quintessence" or the form of a cosmological constant
\cite{wang} \cite{ostr} \cite{cald} \cite{zlat}. Actually, to both these
possibilities is connected a negative pressure, in order to explain the
above quoted observed acceleration: then  we have to require such a
component to have a state equation $p=w\rho$, with $-1< w <0$ (very recent
considerations \cite{perl3} require $-1\leq w
\leq-0.6$ for quintessence, see
also \cite{turn}, \cite{chi1}, \cite{chi2}, \cite{chi3}, for the $x$-matter
scenario); when $w = -1$ we have the cosmological constant scenario
\cite{stein}. \\ In this letter we clarify
 some
questions connected with these different models improving the results
already published \cite{noi1}. In the context of non-minimally coupled
quintessence theories (quintessence is, from this point of view, directly
related with geometry  \cite{perr1}, \cite{noi1}), we give  an exact
treatment of the
 models presenting one of the most commonly used quintessence potential,
i.e.\ the inverse power potential $V=V_{0}Q^{-\alpha}$, with $\alpha>0$
(for example see \cite{peeb} \cite{rat}). We, also, solve exactly the model
connected, in the minimal coupling case, to another type of potential now
present in the literature, i.e.  $V(Q)=V_{0} \sinh(\alpha Q)$, \cite{uri}.
In this scenario we shall discuss a definition (already given in a
different context \cite{noi3}) of quintessence effective cosmological
constant, and then, using it in the cases we are considering, we explain
some of the most ad hoc  $\Lambda$ decaying behaviours considered  in
literature \cite{ov}. We will spend, also, some words on the
($\Lambda$-)fine tuning \cite{zlat} and the cosmic coincidence problems
($\Omega_{m}
\sim
\Omega_{\Lambda}$)
(see for example \cite{stein}), which are present in these scenarios. We
believe that such an exact treatment of the models can put some further
light on the two above quoted problems connected with  these scenarios.
Furthermore we want to mention, in these introductory remarks, that,
concerning the use of $\Lambda$ to explain the supernovae results, doubts
have been presented: for example it has been shown that the inhomogeneity
can be used to explain them \cite{Cel} \cite{noi10}. Together with this,
Riess et al. \cite{Riess} have opened some doubts on the interpretation of
same data stressing the possibility of interpreting them entirely in the
context of a standard open Friedmann-Robertson-Walker model joined with a
reasonable astrophysical evolution model of the white dwarf supernovae
progenitors. Finally we will exhibit a sort of complementarity relation
showing the usual mutually exclusive presence, in these models, of the
quintessence evolving dynamics and of the given (effective quintessence)
cosmological constant evolving dynamics. This result, together with finding
a very well studied $\Lambda$-decaying from our quintessence defined
effective cosmological constant, we believe is the bridge which makes
evident the difference in using a quintessence dominated Universe or a
quintessence defined effective $\Lambda$ dominated Universe. This means
that the last one can be "deduced" from the former one and, in a dynamical
way, it could be connected to an asymptotic really constant $\Lambda$ at
the end (this last aspect is interesting also because we think it can
reintroduce a new attention in the cosmological NoHair theorem \cite{wal}).

We start considering the general (field) action that describes the model we
are going to use:
\begin{equation}
{\cal A} = \int \sqrt{-g} (F (Q) R +
\frac{1}{2}g^{\mu \nu} Q_{, \mu} Q_{,\nu} - V
(Q) + {\cal L}_m)d^4x
\end{equation}
being $(F(Q),V(Q))$ the (generic) functions describing respectively the
coupling and the potential, $R$ is the curvature scalar, ${\displaystyle
\frac{1}{2}g^{\mu\nu}
Q_{, \mu} Q_{, \nu}}$ is the kinetic energy of the quintessence field and
${\cal L}_m$ describes the standard matter content.
 In units $8 \pi G=\hbar= c = 1$ we recover the standard gravity when
 ${\displaystyle F}$ is
equal to -$\frac {1}{2}.$ In the flat Friedman--Robertson--Walker
cosmologies, (1) gives rise to the "point- like" Lagrangian ${\cal L}$ (we
use this expression intending that the field density Lagrangian connected
to (1), because of the cosmological principle, can be considered as defined
in the minisuperspace where the {\cal remnant} two field variables ($a, Q$)
have to be considered functions only of the cosmological time and then
considered as describing a two degrees of freedom model). In this sense
${\cal L}$  is defined on a two dimensional configuration space (actually
we are treating the quintessence field which is not homogeneous, as a
function of the time only)
\begin{equation}
{\cal L} = 6 F a \dot{a}^2 + 6 F' \dot{Q}a^2 \dot{a}+ a^3 p_Q
- D a^{-3(\gamma -1)}
\end{equation}
where $a$ is the scale factor, $p_Q \equiv {\displaystyle \frac {1}{2}
\dot{Q}^2
- V (Q)}$ and $\gamma$ is given by using the standard matter
$p_m = (\gamma -1) \rho_m$ state equation (completely independent of the
second order differential Eqs. connected to (2)). Prime denotes the (total)
derivative with respect to $Q$, dot the same with respect to time.\\ From
Lagrangian (2) we get the same equations which come from the field Eqs.\
derived from (1) in the FRW metric. Variation of $Q$ gives the
Klein--Gordon equation, whereas the Bianchi identity gives rise to
(standard) $\rho_m=D a^{-3\gamma}$, being the constant $D$ given by
$D=\rho_{m_0}a_0^{3\gamma}$ (the expressions  $(.)_0$ denote the value of
the quantity $(.)$ now). We find from (2)
\begin{equation}
H^2 + \frac{\dot{F}}{F} H + \frac{\rho_ Q}{6 F} +
\frac{{\rho}_m}{6F}=0,
\end{equation}
\begin{equation}
2 \frac{\ddot{a}}{a} + H^2 + \frac{\ddot{F}}{F} + 2 H
\frac{\dot{F}}{F} - \frac{1}{2F} P_Q - \frac{1}{2F}P_m =0,
\end{equation}
where $\rho_Q \equiv {\displaystyle \frac {1}{2}
\dot{Q}^2
+ V (Q)}$.
In this presentation the Klein--Gordon equation is derived from
(3), (4); it is interesting that Eq.(3) is precisely $E_{\cal L}
=0$, being $E_{\cal L}$ the (constant) energy associated with ${\cal L}$.
We will consider here only the dust case, i.e. $p_m =0$. According to
Noether theorem
(the Noether symmetry is studied in the quintessence minisuperspace),
we could get a
further information (the possible existence of that symmetry) very useful to exactly
integrate the system (3), (4) as well as to find a form for the two unknowns
$(F(Q),V(Q))$. The study of the existence of this symmetry actually leads to an
infinite set of Lagrangians (2) if the following relation between the two
functions
$F(Q), \, V(Q)$ is satisfied:
\begin{equation}
V = V_0 (F (Q))^{2 p (s)},
\end{equation}
where  ${\displaystyle p = \frac{3 (s+1)}{2s +3}}$, and if $F(Q)$ is of the
form
\begin{equation}
F = F_{0} (s) Q^2,
\end{equation}
with $ F_{0} (s) = \frac{3 s +2}{48 (s+1) (s+2)}$. The parameter $s$ labels
each Lagrangian belonging to the class of infinite Lagrangians (of type
(2)) admitting a Noether symmetry. The value $s=0$ is a permitted value
and gives rise to a model admitting a Noether symmetry, but has to be
treated in a different way (for a complete treatment of this approach see
\cite{noi2}). Requiring $F (Q) < 0$ (attractive gravity) and $V (Q)$ of
inverse power-law type, we get $s
\in (-3/2\, , \, \, -1)$. In Fig. (1)  we plot the two
functions $F_{0} (s)$, $p(s)$:
\begin{figure}[ht] \centering \mbox{ \subfigure[]{\epsfig{figure=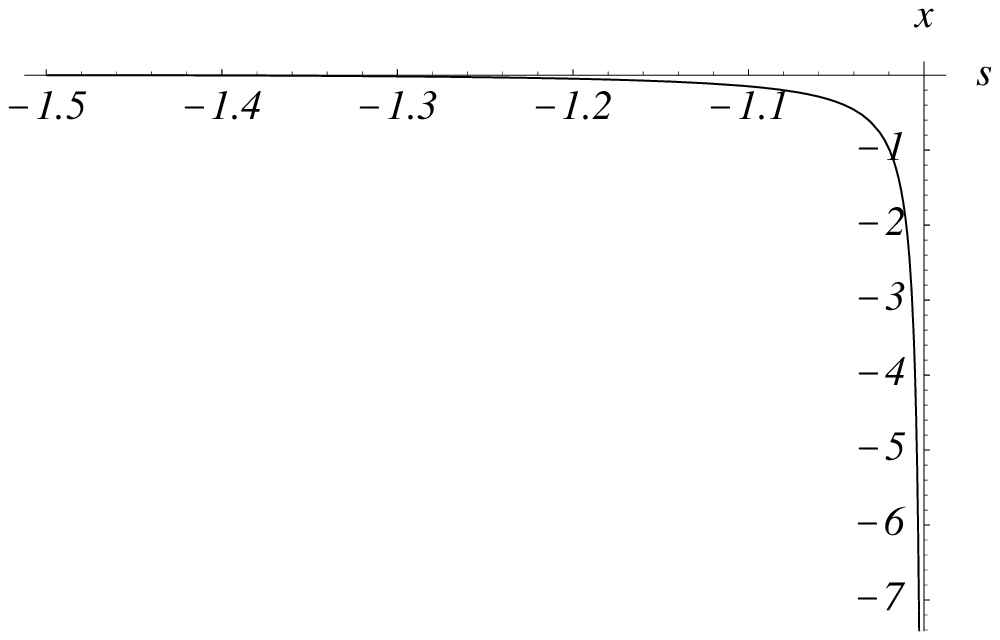,
height=4cm,width=0.4\textwidth,clip=}}
\quad \subfigure[]{\epsfig{figure=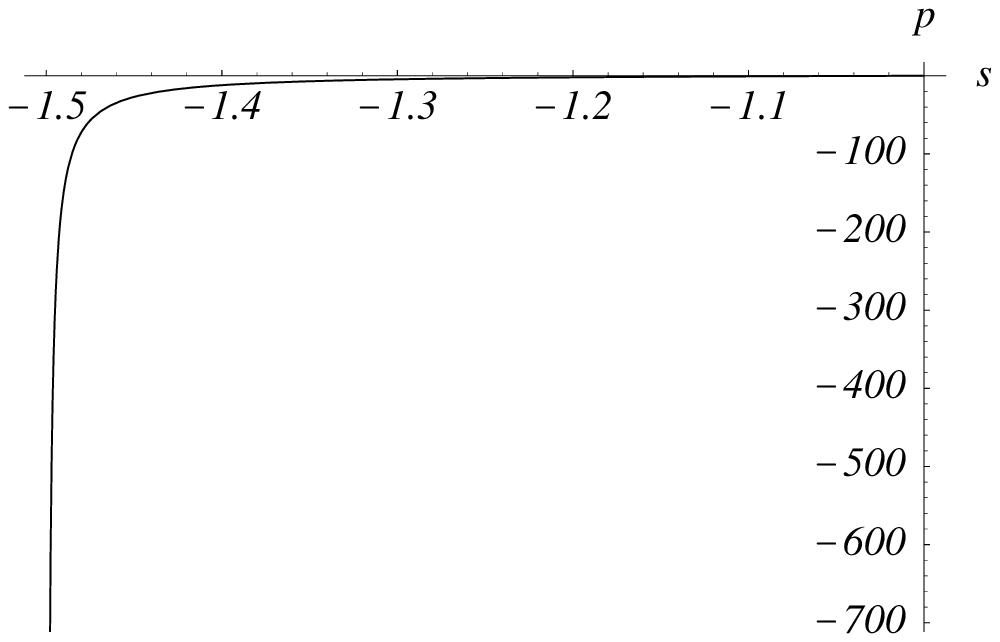,height=4cm,
width=0.4\textwidth,clip=}}}
\caption{{\small In (a) we plot the coefficient of the
function $F(Q)$}. In (b) we show all the  possible exponents for
the inverse power-law
potential: pratically, all exponent values are avalaible.}
\end{figure}\noindent (pratically all the interesting expressions of the
inverse power law potential are available: for example, $ V =
\frac{V_0}{Q^2}$
is relative to the model we pick up fixing the value $s =
-1.257571$). The existence of the Noether symmetry gives a further first
integral of the second order Euler-Lagrangian Eqs. related to (2). Using it
we can exactly solve those equations; their solutions $a(t)$, $Q (t)$ are
\begin{eqnarray}
& & a (t) = \delta_2 (s) [ k_1 t + k_2]^{\frac{s+2}{s+3}} \left\{ k_3 [k_1
t + k_2]^{\frac{s+6}{s+3}} + b_1 t + b_0
\right\}^{\frac{s+1}{2}} \\& & \nonumber \\& & Q (t) = \delta_1 (s)
\left\{[k_1 t + k_2]^{\frac{1}{s+3}}
\left\{ k_3 [k_1 t + k_2]^{\frac{s+6}{s+3}} + b_1 t + b_0
\right\}^{\frac{1}{s}}\right \}^{-\frac{2s +3}{2}} \end{eqnarray}
being
\begin{eqnarray}
& & \delta_1 (s) = \left[ \left( \frac{
\chi(s)}{3}\right)^3 \right]^{\frac{1}{\chi (s)}}\,, ~~~~\chi (s)
= - \frac{6 s}{2 s + 3} ,\, ~~~~ \delta_2 (s)  = \delta_1^{-
\frac{2}{3p(s)}},\nonumber \\
&&b_1 =\frac{-sD}{3\Sigma_0}, \,\,k_1 = \frac{s+3}{s}
\frac{\Sigma_0}{\gamma (s)},\,\,\, k_2 =
\omega_0^{\frac{s+3}{3}},\,\,\,k_3 = - \frac{V_0 (s+3)^2}{3 k_1^2
\gamma(s) (s +6)}\nonumber\\
&& \gamma(s) = \frac{2s +3}{12 (s +1)(s+2)},\nonumber
\end{eqnarray}
where $\Sigma_0$, $\omega_0$, $b_0$ and $b_1$ are the four initial data
(the constant $\Sigma_0$ comes from the  existence of the Noether
symmetry); considering the condition $E_{\cal L}
=0$, we get $3\Sigma_0 b_1 + sD =0$,
representing the only constraint
on the four initial data for the system of the two second order differential
 Eqs.
(7), (8) which, then, become three initial data as usual. From this
constraint we see that neither $\Sigma_0$ nor $b_1$ can be zero because we
are studying models
 with
nonzero standard matter. The constant $D$ comes from the Bianchi identity
for the standard matter as we have already stressed, and it cannot be
considered like an initial datum  for the system (7) and (8) because the
state equation is used. These informations, together with $V_0\neq 0$,
which is quite obvious, tell us that $k_1$ and $k_3$ have to be different
from zero. It is important to stress that $\frac{s+6}{s+3}>1$, for $~~s \in
(-3/2\, ,-1)$;
then, for large $t$ the two functions (7), (8) become $a (t)
= A_0 t^{r(s)}$, $Q
= Q_0 t^{N (s)}$, where
\begin{eqnarray} & & r(s) = \frac{6 + 9s + 2s^2}{s(3 +s)}
~~~~~~~ (> 0,~~~~\mbox{for}~~s \in (-3/2\, , -1)) \,, \\& & \nonumber
\\ & & N (s) = \frac{-(3 + 2s)}{s}~~~~~~~ (> 0, ~~~~\mbox{for}~~s \in
(-3/2\, , -1)) \,,
\end{eqnarray} which we plot in Fig.2.
\begin{figure}[ht]
\centering
\mbox{\subfigure[]{\epsfig{figure=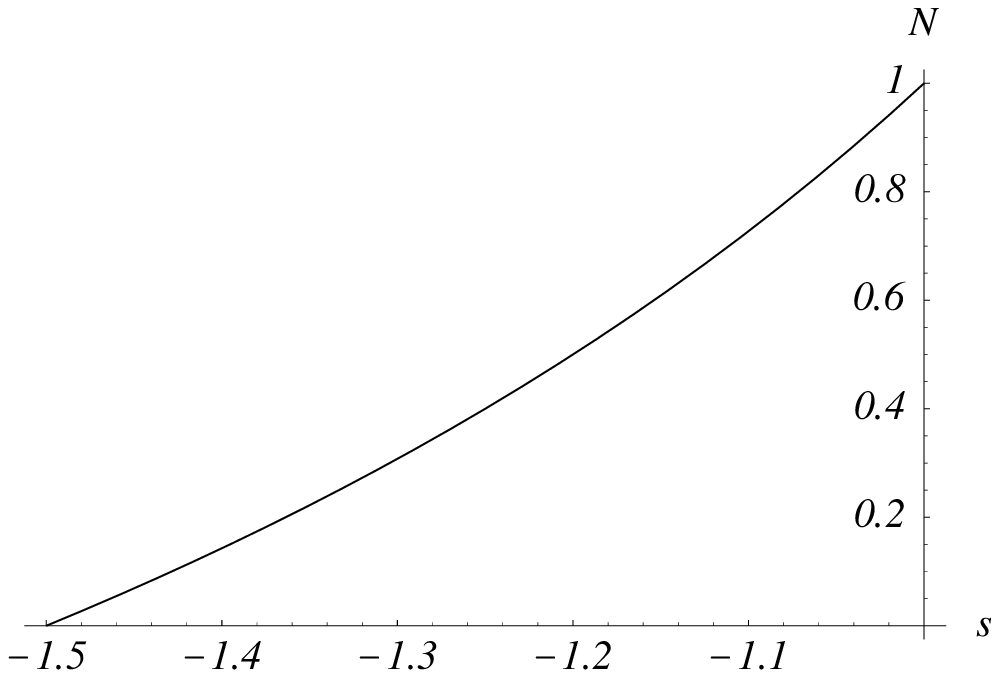,height=4cm,width=0.4\textwidth,
clip=}}\quad\subfigure[]{\epsfig{figure=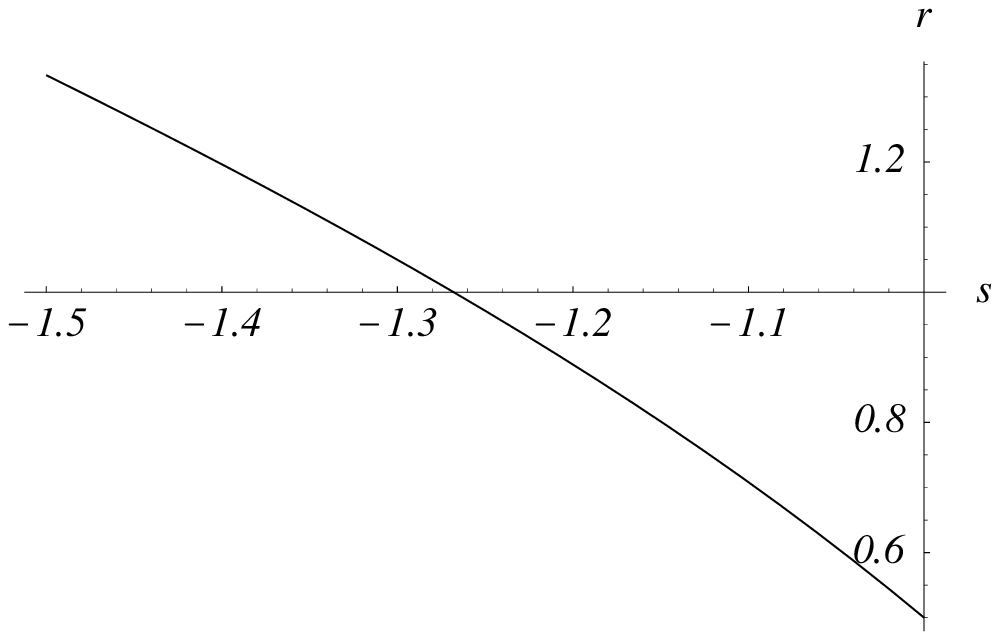,height=4cm,
width=0.4\textwidth,clip=}}}\caption{{\small The plots of the two exponents
we find in the large $t$ behaviour of the two functions $Q$ and $a$
respectively. It is interesting to see the these two functions give rise to
a monotone t-dependence for both the functions we are considering: then the
inverse functions (say $t=t(a)$) are always well defined.}}
\end{figure}
We have indicated with $(A_0, Q_0)$ the two coefficients which come from
(7), (8) for $t \gg 0$: these two constants are parametrized by $s$, that is
they are dependent on the model; they  also depend on  $\Sigma_0$ and
$k_1$. It is important to stress here that this does not give rise to any
real limitation. Actually, we are just facing a typical situation found in
standard cosmology. If we consider, for example, the de Sitter solution
which is given by $a(t)=a_0 e^{{{\sqrt{2\Lambda/3}}t}}$, of course we have
to impose  $a_0>0$ from the very beginning, and this does not imply that
such a behaviour depends on initial data. Actually this is required by
considering an expanding Universe; then the initial data have to belong to
the semiplane $a_0>0$. The same concerns $(A_0, Q_0)$, that is for any of
those data the asymptotic behaviour of $a(t)$ and $Q(t)$ is the same, they
both cannot be zero for all the physical initial data one can choose. We
believe that, in this way, we do not have any (initial data) fine tuning in
our models, this being strictly connected to the control we have on those
data; that's why is so important to have the exact solutions of the
evolution of the cosmological variables. It is also relevant that we find
most of the (large t) solutions spread out in the literature. We see that
$Q (t)$ diverges for any s $\in(-3/2,\,-1)$, the same is true for $a(t)$:
in the same $s$-interval $\dot{Q}$ goes to zero for large $t$. \\ Of
course,  we can fix a special value of $s$ (which identifies the model) in
order to get special time dependences for the scale factor: for example, we
can have
\begin{equation}
a \sim t^{2/3}\,
\end{equation}
for $s = -1.0788$. For this value of $s$, the potential $V (Q)$ is given by
${\displaystyle
\frac{\bar{V}_0}{Q^{0.56}}}$,
 $ (\bar{V}_0 = F_{0} (\bar{s})^{2p(\bar{s})}V_0)$.\\
$Q (t)$ diverges like $t^{0.790}$ (for $s
= - 1.0788)$.
In Table1 are given particular  values of the s-functions which play a
relevant role (actually, we recover almost all of the studied beaviours for
$a(t)$).
 \begin{table}

\begin{center}
\begin{tabular}{c|c|c|c|c|c} \hline
$s$ & $r(s)$ & $1/r(s)$ & $N(s)$ & $1/N(s)$ & $2 p(s)$ \\ \hline
-1.0788 & 0.666  & 1.5001 & -0.780 & 1.280 & -0.280 \\
~ & ~ & ~ & ~ & ~ & ~ \\
-1.1 & 0.708 & 1.412 & -0.727 & -1.375 & -0.75 \\
~ & ~ & ~ & ~ & ~ & ~ \\
-1.2 & 0.888 & 1.125 & -0.5 & 2 & -2. \\
~ & ~ & ~ & ~ & ~ & ~ \\
-1.3 & 1.0497 & 0.952 & -0.307 & 3.25 & -4.5 \\
~ & ~ & ~ & ~ & ~ & ~ \\
-1.4 & 1.196 & 0.8358 & -0.142 & 7 &-0.280 \\
~ & ~ & ~ & ~ & ~ & ~ \\
-1.499 & 1.332 & 0.750 & 0.001 & 749 & -4.5\\ \hline
\end{tabular}
\end{center}
\caption{{\small For different choices of the parameter $s$ we give
the values of the functions describing the  most important behaviours
appearing in each models we consider. It is interesting to stress that for
$s=-1.499$ we get the radiation behaviour for the scale factor.}}
\end{table}
Furthermore we see that, being $\tilde{\rho}_Q
=\displaystyle{\frac{\dot{F}}{F} H }+
\frac{\rho_Q}{6 F}$, and substituting the solutions we found, we
have ${\displaystyle
\tilde{\rho}_Q \sim \frac{1}{t^2} =
\frac{1}{a^{2/r}}}$, in the case $s=-1.0778$-model, we get $r=3/2$,
 and ${\displaystyle \tilde{\rho}_Q \sim \frac{1}{a^3}}$. Then, among the
infinite models under considerations, there is an exactly integrated one
which shows a scaling-type behaviour \cite{lid}, $\tilde{\rho}_Q\sim
\displaystyle{\frac{1}{a^3}}$ scales like $ \rho_m$
(we also  see that there are a large number of models exhibiting $2/r\leq
3$ (tracker behaviour, see \cite{cald}, \cite{zlat}). In TableI we find
also the relative values for the power in the expression of the potential.
\\The treatment reported before requires $s\neq 0$ \cite{noi2}; as we said
the case connected to the value $s=0$ has to be discussed separately. It
can be shown that two subcases are found: there is a Noether symmetry for
the Lagrangian (2) if: (i)$F= K_0 Q^2$, $(K_0 <0)$ and $V = V_0 Q^2$ $(V_0>
0)$; (ii) ${\displaystyle F=-\frac{1}{2}}$ (minimal coupling), and $V
= V_0(A e^{\mu Q}
- Be^{-\mu Q})^2$, with $\mu =\sqrt{3/2}$ \cite {uri}.

It is easy to see that in this last case the starting Lagrangian we have to
consider is ${\cal L}= 3a \dot{a}^2 - a^3(\frac{1}{2}\dot{Q}^2-V(Q))- D$
for the standar dust case. Everything goes like the $s\neq 0$-cases  and we
will not discuss here the problems related to the initial data as well as
the constraint ( Einstein (0, 0) Eq.) imposes on the four initial data. We
will discuss only the minimal coupling case  which splits in two subcases
(for the complete discussion of both cases, see \cite{noi2}). Depending on
the relative signs of the two free parameters $(A,B)$, the general
solutions of FRW Eqs. are:\\ $AB>0$
\\
\begin{eqnarray}
&& a(t) = \left[\frac{\omega_0(\sqrt{AB}t + z_0)^2 -\omega_0^2 \sin^2
(\sqrt{AB} t+ \omega_1 )}{4AB} \right]^{1/3}\\
&&Q(t)=\sqrt{\frac{2}{3}}\ln{\sqrt{\frac{B}{A}}\left[
\frac{(\omega_0 \sqrt{AB} t + z_0) + \omega_0 \sin (\sqrt{AB} t +
\omega_1)}{(\omega_0 \sqrt{AB} t + z_0) - \omega_0 \sin (\sqrt{AB}
t + \omega_1)}\right]}\\ \nonumber
\end{eqnarray}
$AB<0$
\begin{eqnarray}
 a(t)= \left[ \frac{-(\omega_0 \sqrt{-AB} t + z_0)^2 +
\omega_0^2 \sin h^2 (\sqrt{-AB} t + \omega_1)}{4|AB|} \right]^{1/3}\\
 Q (t) =\sqrt{\frac{2}{3}} \ln{\left\{\sqrt{\left|\frac{B}{A}\right|}
 \left[ \frac{(\omega_0 \sqrt{-AB}t + z_0) +
\omega_0 \sin h ( \sqrt{-AB} t + \omega_1)}{-(\omega_0 \sqrt{-AB}
t + z_0) + \omega_0 \sin h (\sqrt{-AB} t + \omega_1)} \right]\right\}}
\end{eqnarray}
where $\omega_0$, $\omega_1$, $z_0$ are the three integration constants (of
course $A$, $B$ and $D$ are different from zero). In both cases,
$Q\rightarrow $ const. for large $t$; in the first case $a(t)$ behaves as
$t^{4/3}$ for small $t$ and as $t^{2/3}$ for large $t$ (self tuning
solution see \cite{lid}), in the  second $a(t)$ has a de Sitter asymptotic
behaviour. Anyway, the asymptotic behaviours are independent of initial
data in the same sense that we have clarified above.
\\ After having exhibited exact solutions for inverse power law potential, and
for some kind of exponential potential, let us go now to discuss a way of
introducing an effective, time dependent, cosmological "constant"
\cite{noi3} (we hope in this way to solve the connected fine tuning problem
\cite{car}). Before presenting our definition of an effective, cosmological
"constant" it is notewhorty to recall that: i) standard $\Lambda$ is
introduced by hands; ii) it determines the (in general asymptotic) time
behaviour of $a(t)$ through the (0, 0)-FRW Eqs., which can be rewritten as:
$\left( H
-
\sqrt{\frac{3
\Lambda}{2}}
\right)
\left( H + \sqrt{\frac{3 \Lambda}{2}} \right) = \varrho_m$. An  expanding
universe requires, asymptotically, ${\displaystyle H =
\sqrt{\frac{3
\Lambda}{2}}}$. If we look at our (3) we see that it is
still possible (in the dust case) to put it in the similar form ${\displaystyle (H -
\Lambda_{\mbox {\small eff}_1} ) ( H + |\Lambda_{\mbox{\small
eff}_2}|)\equiv -\frac{\rho_m}{6F}}=-\frac{\tilde\rho_{{eff.}m}}{3}$, being
\begin{equation}
\Lambda_{\mbox{\small eff}_1} = - \frac{\dot{F}}{2F} +
\sqrt{\left( \frac{\dot{F}}{2F}\right)^2 - \frac{\rho_{
Q}}{6F}} ~~~~~(>0,~~~\mbox{because}~~~F < 0)\, .
\end{equation}
The second root, i.e. $\Lambda_{\mbox{\small eff}_2}$, is less than zero
and does not affect the asymptotic time behaviour of the cosmological
quantities. Two comments on definition (17) are in order: first, it is
completely defined on the quintessence side of the quintessence-tensor
theories we are considering, i.e. it is defined only using $Q$ once we have
$F(Q)$ and $V(Q)$; from this point of view, $\Lambda_{\mbox{\small eff}_1}$
is not introduced by hands but using the same procedure we have mentioned
above, that is using the roots of the (0, 0) Einstein Eqs. We, also, want
to stress that it could be of some interest to study, in the effective
cosmological constant scenario, under which conditions (17) becomes, for $t
\gg 0$, a constant giving back the condition for having a de Sitter
asymptotic behaviour (the complete discussion of how the asymptotic
cosmological NoHair Theorem can be generalized to this case is in
\cite{noi3}). From definition (17) we have that, using the solutions of the
system (3) and (4), $\Lambda_{\mbox{\small eff}_1}$ is a function of $(Q
(t),
\,
\dot{Q} (t))$, and of the
parameters connected with $F (Q)$ and $V (Q)$. It is noteworthy that for $s
\in (-3/2, , -1)$ it is found that $\Lambda_{\mbox{\small eff}_1}$ decays in a
way very well studied in literature, even if in all these discussions the
decaying of $\Lambda_{\mbox{\small eff}}$ is given ad hoc (see the
exhaustive paper by Overduin and Cooperstock \cite{ov}).
\\Using the solutions (7), (8) for $t \gg 0$, we find:
\begin{equation}
\Lambda_{\mbox{\small eff}_1} = \frac{{\cal B}}{a^m}\, , ~~~~~~ m =
m (s) = \frac{1}{r (s)}\, .
\end{equation}
Even if the constant ${\cal B}$ is depending on $s$ and on the initial data
of the problem as well as on the parameter $V_0$, as we have stressed till
now,  the way $\Lambda_{\mbox{\small eff}_1}$ behaves in time does not
depend on the initial data. Then Eq. (18) shows that the form of the
decaying is general, that
 is
$\Lambda_{\mbox{\small eff}_1}$ goes to zero independently of any initial
data. In \cite{ov} the form (18) is given a-priori and then an explanation
of the constant $\cal B$ is not found (in terms of $H$ we have also that
$\Lambda_{\mbox{\small eff}_1} \sim H$). When $a(t)
\sim t^{2/3}$ we have $m = 1.50016$, that is $\Lambda_{\mbox{\small eff}_1}
\sim{\displaystyle \frac{1}{a^{1.5}}}$ compatible with lens statistics,
power spectrum of matter density perturbations \cite{ov}. Also the value $m
=2$ is permitted for $s = -1.0002$.
 In the case $s =0$, with $AB >0$ we
have that ${\displaystyle \Lambda_{\mbox{\small eff}_1} \sim
\frac{\bar{\cal B}}{a^{3/2}}}$,
($\bar{\cal B}$ is the equivalent of $\cal B$  introduced above, in the
case $s=0$, ${Q}$ goes to const.). For $AB <0$ we get the interesting
result $\Lambda_{\mbox{\small eff}_1}
\rightarrow \Lambda (A,B) =
\mbox{const.}$, independently of any initial data and then we
recover a de Sitter asymptotic behaviour for the scale factor (also in this
case ${Q}$  goes to const.).



In summary, in the context of Noether symmetry approach to non-minimally
and minimally coupled quintessence tensor theories of gravity we have
obtained a class of models exhibiting two important types of potentials,
respectively, inverse power law and exponential, and we have exactly solved
it. If we look at the solutions, we can state a property which shows a sort
of complementarity between quintessence and the dynamically defined
$\Lambda_{\mbox{\small eff}}$. Before writing down this property, in Table2
we give the behaviours we have found for $(a (t)$, $Q (t))$, $\dot{Q (t)})$
and for the effective $\Lambda$-term we have introduced.
\begin{table}
\begin{center}
\begin{tabular}{c|c|c} \hline
$s \in (-3/2,~ -1)$ & $s= 0,~ AB > 0$ & $s= 0,~ AB < 0$ \\ \hline$a(t) \sim t^{r(s)}$
& $a(t) \sim t^{2/3}$ & $a(t) \sim
\exp[\Lambda_{\mbox{eff}_1} t]$  \\~ & ~ & ~ \\ $Q(t) \sim \infty$ & $Q(t)
 \sim const.(\neq 0)$ & $Q(t)\sim const.(\neq 0)$
\\ ~ & ~ & ~ \\ $\dot{Q}(t) \sim 0$ & $\dot{Q}(t) \sim 0$ & $\dot{Q}(t)
\sim 0$ \\
 ~ & ~ & ~ \\ $F(Q(t))\sim\infty$ & $F(Q(t))= -1/2$ & $F(Q(t))= -1/2$ \\
  ~ & ~ & ~ \\
  $\dot{F}/F \sim 0$ & $\dot{F}/F =0$ & $\dot{F}/F =0$ \\~ & ~ & ~ \\
  $V(Q(t)) \sim 0$ & $V(Q(t)) \sim 0$ & $V(Q(t)) \sim const. (\neq 0)$ \\
   ~ & ~ & ~ \\
   $\Lambda_{\mbox{eff}_1} \sim 1/a^m$ & $\Lambda_{\mbox{eff}_1} \sim\frac{1}{a^{\frac{3}{2}}}$ & $\Lambda_{\mbox{eff}_1} \sim const. (\neq 0)$ \\ 
\hline
\end{tabular}
\end{center}
\caption{{\small The $t\gg0$ behaviour of all the physical
quantities introduced. It is noteworthy that in all the cases the potential
rolls down to its minimum. In the $AB>0$ case its minimum is zero, whereas
in the $AB>0$ case the minimum is different from zero, and then we recover
a real cosmological constant. We see also that the kinetical-Q energy has
no asymptotical role}}
\end{table}
The relation we propose is the following:
\begin{equation}
\Lambda_{\mbox{\small eff}_1} f(Q)) = \mbox{cost.}\,
, ~~~~~~ t\gg 0,
\end{equation}
where $f$ depends not only on the $Q$-field but is determined also by
 $s$ (for example is given by $Q^{\frac{1}{N(s)}}$
in the cases $s$ $\in(-3/2,-1)$, i.e. it is determined by the model we
choose. Relation (19) is true, in general, if we assume the esistence of
Noether symmetries and that $\frac{\dot{F}}{{F}}=\dot{G}_{\mbox{\small
eff}}
/ G_{\mbox{\small eff}}
\stackrel{ t \gg 0}{\longrightarrow} 0$ (see \cite{noi2} for a better
understanding of the hypothesis on
$\frac{\dot{F}}{{F}}=\dot{G}_{\mbox{\small eff}}
/ G_{\mbox{\small eff}}
\stackrel{ t \gg 0}{\longrightarrow} 0$).
We see that (19) gives rise to the already mentioned complementarity
between the $\Lambda$-term and the (divergent or convergent to a constant)
quintessence content in the universe. From this point of view we do not
have that dark energy, or quintessence, and cosmological constant are
completely different forms of energy: first, because they are different
forms of the same (quintessence) energy; in fact, we derive the
$\Lambda$-term from the presence of quintessence. Furthermore, they are
complementary in the sense given by (19) (actually, the quintessence
kinetic energy has no asymptotic role in all the cases we have presented),
from which we deduce that the dominant ingredient is connected with the
large $t$ behaviour of the Q-field and with the form of the potential
minimum. When this minimum is zero we have that the Q-field dominates, when
this minimum is non zero we have the asymptotic dominance of the effective
cosmological "constant". In this paper we do not discuss the way $\Lambda$,
as well the solutions we have found, depends on the potential parameter; to
this purpose see our \cite{noi1}. From Table2 it is clear that the two
possible evolutions of the scale factor are quite different, then the
expansion history of the Universe is different: its acceleration and its
age will strongly depend on the dominance of the Q-matter or of the
effective $\Lambda$-Q. In all cases the potential rolls down to its minimum
and then, as we have stressed, the value of this minimum plays a very
important role: in the cases we have presented, when this minimunm is
different from zero, we have an asymptotic cosmological constant (de Sitter
behaviour for the scale factor); in the cases this minimum is zero, we do
not recover any cosmological constant (power expansion of the scale
factor). More precisely, we have that in the (Noether) nonminimal coupling
case the minimum is always zero, whereas the coupling is divergent; then we
get that $\dot{G}_{\mbox{\small eff}}
/ G_{\mbox{\small eff}}
\stackrel{ t \gg 0}{\longrightarrow} 0$
and that $\Lambda$ decays. In the (Noether) minimal coupling case,
 when $AB>0$, the potential minimum is zero, whereas in the $AB<0$ case the
minimum is different from zero, and then we recover an asymptotical true
cosmological constant. In \cite{noi1} we discussed in more details the
possible values of the parameter $w$, finding a range of values, at least
for the case $s\in (-3/2,
-1)$,
coherent with the currently used range of values as given in the
introductory remarks. We have to say that, concerning $w$, compared to what
is found in the literature, we have followed here a different approach: we
have not introduced this new parameter but we have reconstructed it using
the knowledege of the exact solutions: more precisely the state equation
$p_{Q}= p_{Q}(\varrho)$ has been found knowing $p_{Q}= (t(\rho_{Q}))$, that
is using $a(t)$ and $Q(t)$ for $t \gg0$. Concerning the large scale structure
connected with this approach, see our paper \cite{noi4}. We believe that,
through the knowledge of the exact solutions we have presented, it is
possible to have a better control of the role of initial data; actually,
all along the paper we have shown that using our approach we do not have
that problem. Furthermore, using our approach, we hope to have clarified,
among the reported other features, some aspects of quintessence models:
more precisely we have given  relation (5), we believe interesting, between
the coupling and the potential as well as the complementarity relation (19)
which holds for the types of couplings and potentials we have studied.\\
\noindent Acknowledgments\\
We like to thank S. Capozziello, C. Rubano and P. Scudellaro for all the
discussions we had together on this topic and C. Baccigalupi and F.
Perrotta for the suggestions they gave on the manuscript.

\end{document}